\documentclass[acmsmall]{acmart}

\AtBeginDocument{%
  }
\setcopyright{none}
\copyrightyear{2025}
\acmYear{2025}
\acmConference[PEARC'25]{Practice and Experience in Advanced Research Computing}{July 20--24,2025}{Columbus, Ohio}
\begin{document}

\title{Using the Open Science Data Federation for data distribution:  Big Bear Solar Observatory use case}

\author{Sydney Montiel}
\email{smontielm2001@alumno.ipn.mx}
\affiliation{
  \institution{Instituto Politécnico Nacional}
  \city{Gustavo A. Madero}
  \state{CDMX}
  \country{México}
}

\author{Alexsandra Guadarrama}
\email{alexaguri07@gmail.com}
\affiliation{
  \institution{Instituto México de Baja California}
  \city{Tijuana}
   \state{Baja California}
  \country{México}
}

\author{Frank Würthwein}
\email{fwuerthwein@ucsd.edu}
\affiliation{
  \institution{University of California San Diego}
  \city{San Diego}
  \country{USA}
}

\author{Fabio Andrijauskas}
\email{fadnrijauskas@ucsd.edu}
\authornotemark[1]
\affiliation{
  \institution{University of California San Diego}
  \city{San Diego}
  \country{USA}
}

\renewcommand{\shortauthors}{Andrijauskas et al.}

\begin{abstract}
The growing demand for extensive data processing is now a standard in many scientific fields. Efficiently distributing data to processing sites and enabling seamless sharing has become crucial. The Open Science Data Federation (OSDF) builds on the success of the StashCache project to establish a global data distribution network. By expanding StashCache, OSDF integrates additional data origins and caches, enhancing accessibility and performance (20 origins and 30 caches), new access methods, and monitoring and accounting mechanisms. Additionally, the OSDF has become essential to the US national cyber-infrastructure landscape due to the sharing requirements of recent NSF solicitations. One use case for the OSDF is the data access to the Big Bear Solar Observatory (BBSO). Integrating the BBSO data into the OSDF provided standard and reliable data access. Moreover, the OSDF caches provide local data worldwide. Using the OSDF and the BBSO data, creating a pipeline to apply image processing techniques to all images from BBSO anywhere on the planet was possible.
\end{abstract}



\begin{teaserfigure}
\centering
  \includegraphics[width=10cm]{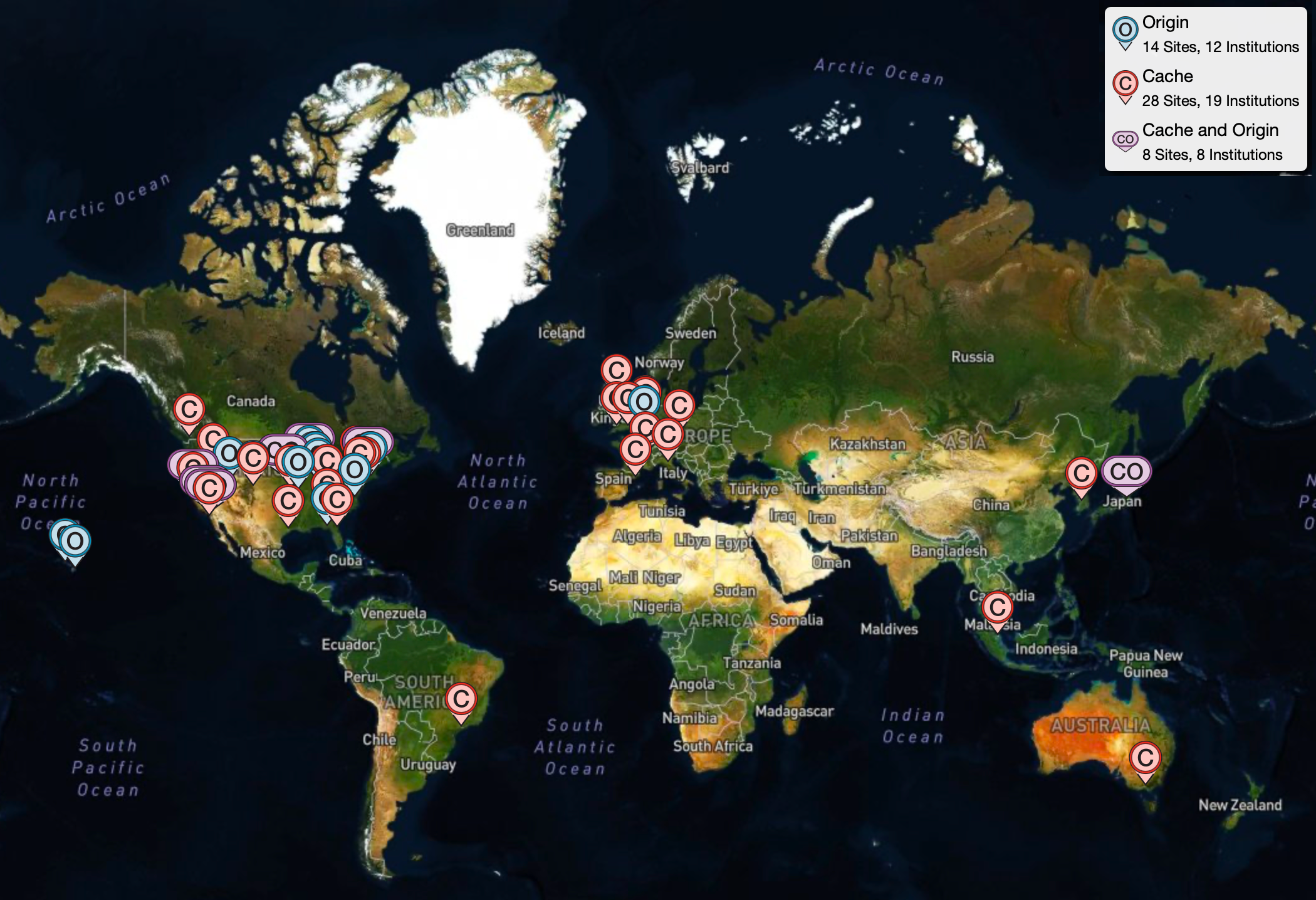}
  \caption{Map featuring the locations of current OSDF architectural components (https://osg-htc.org/services/osdf).}
  \label{fig:teaser}
\end{teaserfigure}

\maketitle

\section{Introduction}

Data sharing has become one of the most complex challenges researchers face in a technological age. It must be stored, managed, processed, and analyzed to generate scientific advances. Therefore, current research must satisfy processing speed and memory requirements. Furthermore, such data is shared by geographically distributed users, which increases network traffic due to access paths \cite{Deng_2023}. Taking this problem as a starting point, the development of the Open Science Data Federation (OSDF) emerged. Based on the Open Science Grid (OSG) project, StashCache, the OSDF is a data access framework that provides infrastructure and tools for accessing data globally, following the notion of "Any Data, Anytime, Anywhere," making data access more efficient for users. It supports large amounts of data from several independent experiments and funded projects by the National Science Foundation (NSF). 

This paper aims to show an OSDF application that can access data from the Big Bear Solar Observatory (BBSO) and efficiently perform image processing for the user.

\section{Open Science Data Federation}

OSDF builds upon the OSG project, StashCache \cite{10.1145/3332186.3332212,Fajardo_2018}, creating a robust research environment in collaboration with OSG. They provide users with high computational throughput and efficient data access \cite{schultz2023icecube}. The OSDF environment continually enhances its tools and capacity to support growth and demand while providing reliable data delivery to compute sites. Additionally, it is essential to empower staff to identify and resolve potential issues early. To facilitate data access for computational workflows running on distributed computing infrastructure, the OSG leverages the Open Science Data Federation (OSDF) \cite{10.1145/3332186.3332212}. At the core of this data delivery system are "origins," "caches," and "redirectors," all implemented using the XRootD \cite{xrootd,xrootd-paper} software framework, which enables low-latency and scalable data access. Figure \ref{fig:teaser} shows all the OSDF host locations. The Pelican Project created a new layer for the OSDF \cite{pelicanproject}, providing centralized services for registration, cache selection, and monitoring.

Origins serve as the backend storage for project data. Within OSDF, an origin is an XRootD/Pelican configuration that provides access to storage through a data transfer node that mounts a project directory. Multiple origins form a hierarchical structure that connects to a redirector, which then communicates with the cache network. Applications typically access OSDF through the nearest cache to their computing site, with proximity determined using GeoIP \cite{10.1145/3332186.3332212}.

A containerized approach on a federated Kubernetes infrastructure is ideal for deploying and managing OSDF software and services, whether at origins or caches. This federated model enables effective monitoring of data access and troubleshooting of issues.

\section{BBSO data processing}

The Sun generates energy by fusing hydrogen atoms into helium. Various events can occur throughout this process, including solar storms, which are particularly interesting due to their potential impact on telecommunications. These storms can disrupt GPS signals, alter or destroy satellite trajectories, and cause electricity transmission issues. For instance, geomagnetically induced currents from solar storms can overload power grids. However, such catastrophic consequences can be mitigated if solar storms are detected in time, allowing for preventive measures such as shutting down satellites or reducing power flow in transmission lines.

\begin{figure}[ht]
         \centering
         \includegraphics[width=8cm]{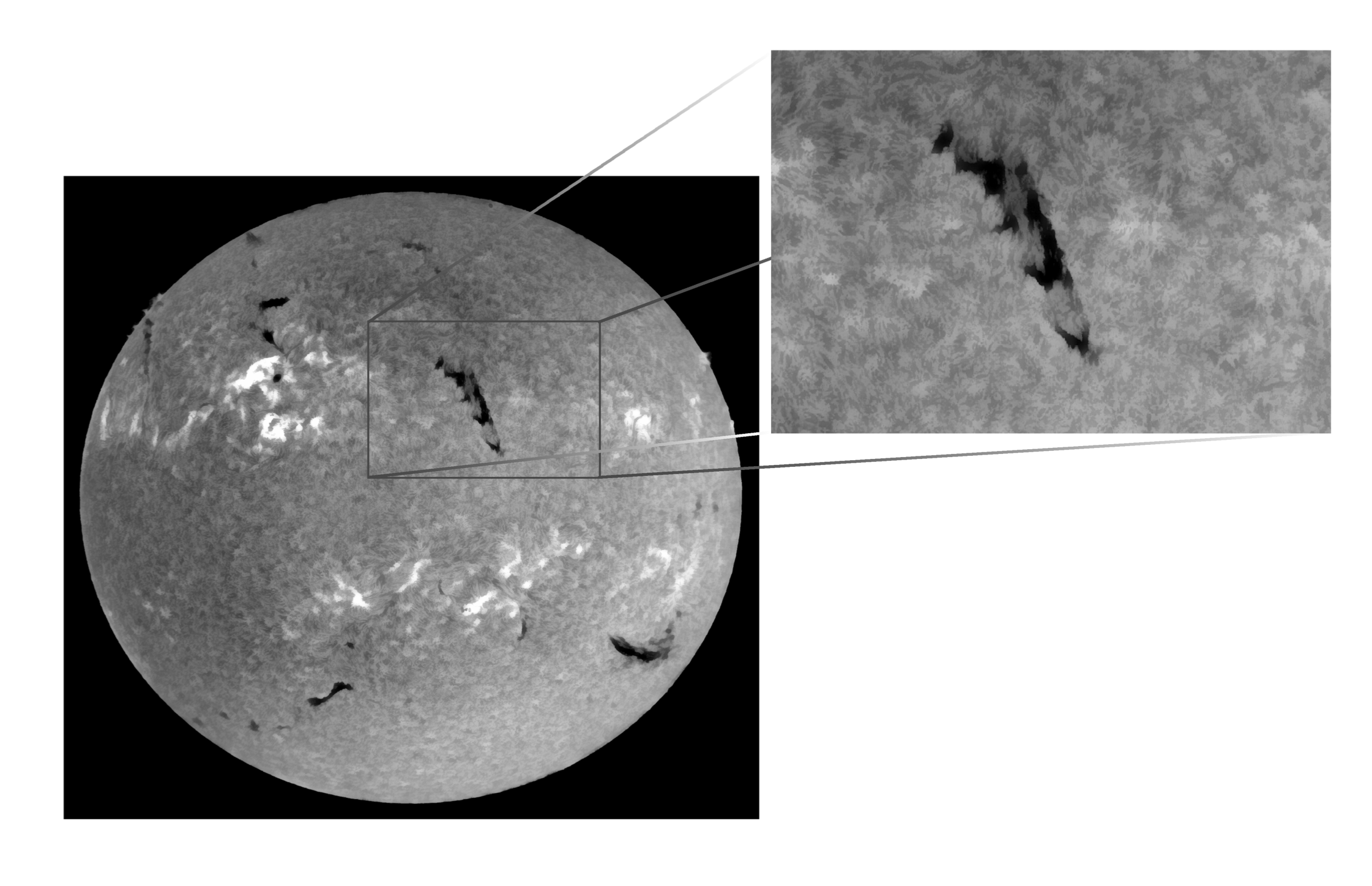}
         \caption{Sun image processed by the diﬀusion filter
with a manually detached filament \cite{10.1145/2286976.2286987}.}
         \label{sun}
\end{figure}

Solar storms exhibit distinct characteristics, one of which is the emergence of filaments on the Sun's surface. These filaments, visible in H$\alpha$ images, consist of dense plasma held in place by magnetic fields. They appear darker than the surrounding solar surface due to their lower temperature. Figure \ref{sun} illustrates a solar storm event featuring a filament, where the dark ribbons on the Sun represent these structures \cite{10.1145/2286976.2286987}.

\begin{figure}[ht]
         \centering
         \includegraphics[width=10cm]{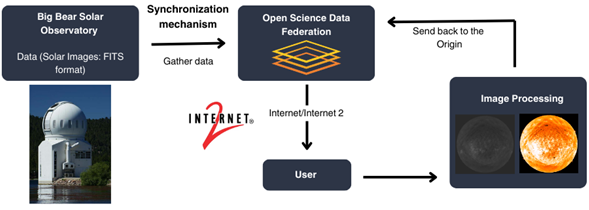}
         \caption{Data flowchart of the gathered images from the BBSO to perform processing.}
         \label{osdfn1}
\end{figure}

Figure \ref{osdfn1} illustrates the flow of solar image data from the Big Bear Solar Observatory through the Open Science Data Federation (OSDF) to users and back. The process begins with the observatory collecting solar images in FITS format. These images are then gathered through a synchronization mechanism and transferred to OSDF, an intermediary data storage and distribution system.

Once in OSDF, the data is transmitted via Internet/Internet2 to users, who access and process the images. The image processing stage enhances the solar images, as shown in Figure \ref{osdfn1}. After processing, the refined data is returned to the origin, completing the cycle. This workflow ensures efficient handling, distribution, and enhancement of solar observational data.

An example of a result produced by the diﬀusion filter is
depicted in Figure \ref{osdfn1}. We can observe that the image is more homogeneous than the original images, but its borders are well-defined. The other steps are threshold calculation, filament extraction, and labeling. The OSDF was used to get all the required images to run this detection. 

\newpage
\section{Conclusion}

The Open Science Data Framework is crucial for sharing scientific data across the United States, featuring caches and origins located worldwide. This capability is vital for high-throughput and high-performance computing. With the addition of over 10 new caches and origins, we can deliver more data to a larger number of users. The BBSO use case demonstrates that it is possible to utilize the OSDF to establish an efficient processing pipeline.

\bibliographystyle{ACM-Reference-Format}
\bibliography{basebib}

@inproceedings{10.1145/3332186.3332212,
author = {Weitzel, Derek and Zvada, Marian and Vukotic, Ilija and Gardner, Rob and Bockelman, Brian and Rynge, Mats and Hernandez, Edgar Fajardo and Lin, Brian and Selmeci, M\'{a}ty\'{a}s},
title = {StashCache: A Distributed Caching Federation for the Open Science Grid},
year = {2019},
isbn = {9781450372275},
publisher = {Association for Computing Machinery},
address = {New York, NY, USA},
url = {https://doi.org/10.1145/3332186.3332212},
doi = {10.1145/3332186.3332212},
booktitle = {Proceedings of the Practice and Experience in Advanced Research Computing on Rise of the Machines (Learning)},
articleno = {58},
numpages = {7},
location = {Chicago, IL, USA},
series = {PEARC '19}
}

@misc{schultz2023icecube,
      title={IceCube experience using XRootD-based Origins with GPU workflows in PNRP}, 
      author={David Schultz and Igor Sfiligoi and Benedikt Riedel and Fabio Andrijauskas and Derek Weitzel and Frank Würthwein},
      year={2023},
      eprint={2308.07999},
      archivePrefix={arXiv},
      primaryClass={physics.comp-ph}
}

@article{Fajardo_2018,
doi = {10.1088/1742-6596/1085/3/032025},
url = {https://dx.doi.org/10.1088/1742-6596/1085/3/032025},
year = {2018},
month = {sep},
publisher = {IOP Publishing},
volume = {1085},
number = {3},
pages = {032025},
author = {E Fajardo and A Tadel and M Tadel and B Steer and T Martin and F Würthwein},
title = {A federated Xrootd cache},
journal = {Journal of Physics: Conference Series}
}

@report{xrootd-paper,
  title={XROOTD-A Highly scalable architecture for data access},
  author={Dorigo, Alvise and Elmer, Peter and Furano, Fabrizio and Hanushevsky, Andrew},
  journal={WSEAS Transactions on Computers},
  volume={1},
  number={4.3},
  pages={348--353},
  year={2005}
}

@misc{xrootd,
  title =    "XRootD",
  URL =      "https://xrootd.slac.stanford.edu",
  year = {2024},
note = {[Accessed 20-Apr-2024]},
}

@inproceedings{10.1145/2286976.2286987,
author = {Andrijauskas, Fabio and Gradvohl, Andr\'{e} Leon Sampaio},
title = {Solar filaments detection using parallel programming in hybrid architectures},
year = {2012},
isbn = {9781450313384},
publisher = {Association for Computing Machinery},
address = {New York, NY, USA},
url = {https://doi.org/10.1145/2286976.2286987},
doi = {10.1145/2286976.2286987},
booktitle = {Proceedings of the 2012 Workshop on High-Performance Computing for Astronomy Date},
pages = {41–48},
numpages = {8},
location = {Delft, The Netherlands},
series = {Astro-HPC '12}
}

@misc{pelicanproject,
	author = {},
	title = {{P}elican {P}latform},
	howpublished = {\url{https://pelicanplatform.org/}},
	year = {2024},
	note = {[Accessed 4-Jun-2024]},
}

@inproceedings{Deng_2023, series={HPDC ’23},
   title={Analyzing Transatlantic Network Traffic over Scientific Data Caches},
   url={http://dx.doi.org/10.1145/3589012.3594897},
   DOI={10.1145/3589012.3594897},
   booktitle={Proceedings of the 2023 on Systems and Network Telemetry and Analytics},
   publisher={ACM},
   author={Deng, Ziyue and Sim, Alex and Wu, Kesheng and Guok, Chin and Hazen, Damian and Monga, Inder and Andrijauskas, Fabio and Würthwein, Frank and Weitzel, Derek},
   year={2023},
   month=jul, pages={19–22},
   collection={HPDC ’23} }

\begin{acks}

This work was supported in part by National Science Foundation (NSF) awards \#1836650, CNS-1730158, ACI-1540112, ACI-1541349, OAC-1826967, OAC-2030508, OAC-2112167, CNS-2100237, CNS-2120019, PHY-2323298, the University of California Office of the President, and the University of California San Diego's California Institute for Telecommunications and Information Technology/Qualcomm Institute. Thanks to CENIC for the 100Gbps networks.

\end{acks}

\end{document}